\documentclass[reprint,twocolumn,superscriptaddress,longbibliography]{revtex4-2}

\usepackage[hidelinks]{hyperref} 
\usepackage{graphicx}
\usepackage{amsmath}
\usepackage{amssymb}
\usepackage{siunitx}
\usepackage{braket}

\DeclareGraphicsExtensions{.png,.jpg,.eps}
\usepackage{xcolor}
\hypersetup{
    colorlinks,
    linkcolor={red!50!black},
    citecolor={blue!80!black},
    urlcolor={blue!80!black}
}

\renewcommand{\bm}{\boldsymbol}

\def\m1{{^{-1}}}

\begin{document}

\author{Bastian Zinkl}
\affiliation{Institute for Theoretical Physics, ETH Zurich, 8093 Zurich, Switzerland}

\author{Keita Hamamoto}
\affiliation{Department of Applied Physics, The University of Tokyo, Tokyo 113-8656, Japan}

\author{Manfred Sigrist}
\affiliation{Institute for Theoretical Physics, ETH Zurich, 8093 Zurich, Switzerland}

\title{Symmetry conditions for the superconducting diode effect in chiral superconductors}

\begin{abstract}
We analyze the presence of non-reciprocal critical currents, the so-called superconducting diode effect, in chiral superconductors within a generalized Ginzburg-Landau framework. After deriving its key symmetry conditions we illustrate the basic mechanism for two examples, the critical current in a thin film and a Josephson junction. The appearance of spontaneous edge currents and the energy bias for the formation of Josephson vortices play an essential part in establishing a splitting of the critical currents running in opposite directions. Eventually, this allows us to interpret a superconducting diode effect observed in the 3-Kelvin phase of Sr$_2$RuO$_4$ as evidence for spontaneously broken time-reversal symmetry in the superconducting phase.

\end{abstract}

\date{\today}

\maketitle



\section{Introduction}

Experimentally establishing unconventional superconductivity with chiral Cooper pairing unambiguously in a given material remains a notoriously difficult task, despite numerous possible symmetry constraints leading to, in principle, decisive selection rules~\cite{sigrist1991phenomenological, Grinenko2021comm}. Chiral order parameters break time-reversal symmetry and parity for the reflection with respect to a plane containing the chiral axis. This allows the detection through the polar Kerr effect. Additionally, the chiral state has intrinsic magnetic properties that manifest themselves in spontaneous supercurrents at surfaces, impurities or domain walls, which could be observed by zero-field $\mu$SR or by probing the local magnetic fields through scanning SQUID or Hall microscopes~\cite{Volovik1988, Goryo1999, maeno2001maeno}. These features have been tested for some superconductors which have been identified as candidates for chiral superconductivity, such as Sr$_2$RuO$_4$~\cite{luke1998time, xia2006high}, URu$_2$Si$_2$~\cite{Schemm2015} or UPt$_3$~\cite{Luke1993, Schemm190}, to name the most important ones. 
For all three materials the polar Kerr effect suggests a chiral superconducting phase. For UPt$_3$ as well as Sr$_2$RuO$_4$ the $\mu$SR zero-field relaxation rate indicates intrinsic magnetism in the superconducting phase compatible with a chiral phase. However, the extensive search for surface currents in samples of Sr$_2$RuO$_4$ has so far only led to negative results~\cite{Kirtley2007, Hicks2010}. 

In the present study we would like to introduce another feature of chiral Cooper pairing, which provides also an alternative way of detecting it. Our proposal is connected to the so-called superconducting diode effect, which occurs under certain circumstances if the critical current of a superconductor is different for opposite current directions. Broken time-reversal and inversion symmetry are the general conditions necessary in order to observe non-reciprocal phenomena of this kind. Recently, this effect has been discussed for non-centrosymmetric superconductors, i.e. materials without inversion centres, where time-reversal symmetry is removed, for instance, by a magnetic field generating the non-reciprocal behavior of the critical current~\cite{ando2020diode, daido2021intrinsic, yuan2021supercurrent}.

Before illustrating this effect for two examples, we first examine the properties of the critical current in a two-dimensional chiral superconductor with a strip-like shape of limited width in $y$- and infinite length in $x$-direction, as Fig.~\ref{fig:film_sketch} displays. Under what conditions is the critical current $ I_c $ of this device different for currents running in positive and negative directions along the strip? We denote the absolute values of these two critical currents by $I_c^{\pm}$ and introduce three symmetry operations, which change the sign of the chirality of the superconducting state. For illustration purposes we assume that the orbital part of the chiral state is of the type $ \psi_{\chi} (\bm{k})  = k_x + i \chi k_y$, where $\chi=\pm 1$ indicates the sign of the chirality with chiral axis along $z$, i.e. out-of-plane. The time-reversal symmetry operation $ {\cal T} $ and the two mirror operations $ \sigma_x $ and $ \sigma_y $ act as
\begin{gather}
{\cal T} \psi_{\pm} (\bm{k})  = e^{i\alpha_T} \psi_{\mp} (\bm{k}), \label{T-psi}\\ 
\sigma_x \psi_{\pm} (\bm{k})  =  e^{i \alpha_x} \psi_{\mp} (\bm{k}),~\sigma_y \psi_{\pm} (\bm{k})  =  e^{i \alpha_y} \psi_{\mp} (\bm{k}), \label{sigma-psi}
\end{gather} 
with 
\begin{gather}
	\sigma_x \begin{pmatrix} x \\ y \end{pmatrix} = \begin{pmatrix} -x \\ y \end{pmatrix},~ \sigma_y \begin{pmatrix} x \\ y \end{pmatrix} = \begin{pmatrix} x \\ -y \end{pmatrix} 
\end{gather}
and in our case $\alpha_x=\pi$ and $\alpha_{\cal T} = \alpha_y=0$.

Now we apply these symmetry operations to the critical current $ I_{c} $ in both directions of $x$. Both $ {\cal T} $ and $ \sigma_x $ reverse the current. We thus find the following relations for the critical current: $ {\cal T} I_c^{\pm} = I_c^{\mp} $ and $ \sigma_x I_c^{\pm} = I_c^{\mp}  $, while 
$ \sigma_y I_c^{\pm} = I_c^{\pm}  $. Combining these bulk symmetry operations so as to conserve chirality leads to
\begin{equation}
{\cal T} \sigma_x I_c^{\pm} = I_c^{\pm},
\end{equation}
which does not affect the values of two critical currents in contrast to
\begin{equation}
{\cal T} \sigma_y I_c^{\pm} = I_c^{\mp}
\end{equation}
and 
 \begin{equation}
\sigma_x \sigma_y I_c^{\pm} = I_c^{\mp}.
\end{equation}
Note that $ \sigma_x \sigma_y = C_2 $ corresponds to a rotation by 180$^\circ$ around the $z$-axis passing through a point on the middle line of the strip. In order to 
allow for the diode effect, $ I_c^+ \neq I_c^- $, both operations, $ {\cal T} \sigma_y $ and $ \sigma_x \sigma_y $, should not be symmetries of the strip. Evidently, the relevant condition is an asymmetry with respect to the reflection $ \sigma_y $. On the other hand, the lack of symmetry under $ \sigma_x $ alone would not be sufficient. 

The appearance of a splitting between $ I_c^+ $ and $ I_c^- $ attributed to spontaneously broken time-reversal symmetry has been reported for the so-called 3-Kelvin (3-K) phase of Sr$_2$RuO$_4$. The 3-K phase appears in eutectic samples of Sr$_2$RuO$_4$ with excess Ru, which segregates into $\mu$m-sized metallic inclusions whose interfaces with the bulk material presumably act as a nucleation region for a superconducting precursor around 3~K above the bulk transition temperature at $ T_c \approx 1.5$ K \cite{maeno19983K, sigrist20013K}. This leads to an inhomogeneous network of weakly overlapping superconducting puddles, which upon lowering the temperature from 3 towards 1.5 K evolves from the initial nucleation through a second time-reversal-symmetry-breaking transition at around $ T^* \approx 2.3 $~K \cite{hooper2004split, kawamura2005tunneling, kaneyasu20103K}. The transition at $ T^* $ is associated with the onset of a weak but readily detectable difference between $ I_c^+ $ and $ I_c^- $ \cite{hooper2004split}. Apart from that, this phase transition manifests itself additionally in the appearence of a zero-bias anomaly in the quasiparticle tunnelling spectroscopy \cite{kawamura2005tunneling, kaneyasu20193K}. 

In the following, we will consider two model cases to illustrate the conditions for the superconducting diode effect. The first is an analogue of the strip geometry, a superconducting thin film where we analyze the maximal supercurrent by means of a generalized Ginzburg-Landau formulation for a chiral superconductor in a tetragonal crystal symmetry. The second example is a Josephson junction between two chiral superconductors, for which we derive an effective theory for the Josephson phase. In both cases we will demonstrate that the absence of a mirror symmetry ($\sigma_y $) is essential for the split. Furthermore, we show that this discussion provides a possible explanation for the experimental observations in the 3-K phase of Sr$_2$RuO$_4$. Our analysis also highlights under which conditions non-chiral superconducting phases exhibit the same effect.

\section{Critical current of a thin film} \label{sec:film}

\begin{figure}[t!]
 \centering
	\includegraphics[width=0.8\columnwidth]{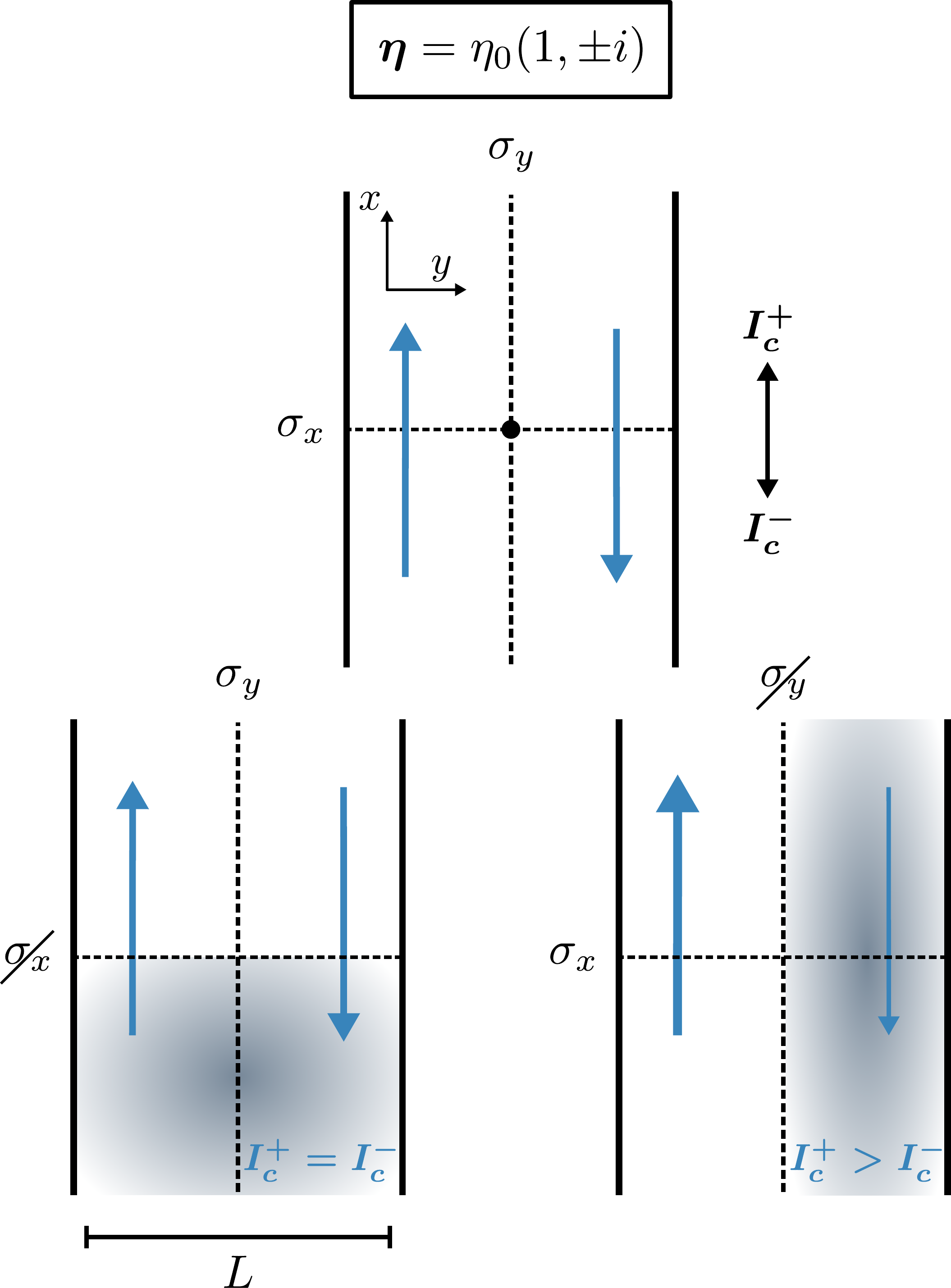}
\caption{Schematic picture of a thin film, which is parallel to the $x$-axis and homogeneous in the $z$-direction. The superconducting state described by the order parameter $\bm{\eta}$ breaks time-reversal symmetry. The plots at the bottom represent cases where additional mirror symmetries are broken.
}
\label{fig:film_sketch}
\end{figure}

In our first example we consider a chiral superconductor shaped as a thin film, which has a narrow width $ L $ along the $y$-direction and extends along $ x$- and $ z$-directions with $ L_x, L_z \gg L $. Thus, the cross-section in the $x$-$y$-plane is the same as shown in  Fig.~\ref{fig:film_sketch}. Assuming that its thickness $L$ is much shorter than the London penetration depth $\lambda_L$ we avoid screening effects. Nevertheless, it will be important to account for the influence of boundary effects on the spatial variation of the order parameter in our discussion. An exemplary superconducting state, incorporating the essential features of the non-reciprocal behavior of $ I_c $ along $x$-direction, is a chiral $p$-wave state given by $ \bm{d}_{\pm} (\bm{k}) = \hat{\bm{z}} \eta_0 (k_x \pm i k_y) $. This spin-triplet pairing state has its chiral axis parallel to the $z$-axis with an equal-spin configuration in the $x$-$y$-plane~\cite{sigrist1991phenomenological}. An important feature for the following discussion is the pair breaking effect 
of the film boundaries in $y$-direction, which we will analyze with the corresponding Ginzburg-Landau theory. 

\subsection{Ginzburg-Landau theory of a chiral superconductor}
The Ginzburg-Landau functional of a chiral $p$-wave superconductor is formulated using the two-component order parameter $\bm{\eta} = (\eta_x, \eta_y)$ $[\bm{d} (\bm{k}) = \hat{\bm{z}} (\eta_x k_x + \eta_y k_y)]$. Assuming an underlying tetragonal point group $ D_{4h} $ the general free energy can be written as
\begin{align}
	&F[\eta_x, \eta_y] = L_z \iint dx dy \left[ f + \frac{1}{8\pi} \left( \bm{\nabla} \times \bm{A} \right)_z^2 \right]  , \label{eqn:pwaveExp}
\end{align}
with 
\begin{align}
	&f = a(T) |\bm{\eta}|^2 + b_1 |\bm{\eta}|^4 + \frac{b_2}{2} \left( \eta_{x}^{\ast 2} \eta_{y}^{ 2} + \eta_{y}^{\ast 2} \eta_{x}^{2}  \right) + b_3 |\eta_{x}|^2 |\eta_{y}|^2  \nonumber \\[1mm] 
	&+ K_1 \left( |\Pi_x\eta_x|^2  + |\Pi_y \eta_y|^2 \right) + K_2 \left( |\Pi_x\eta_y|^2  + |\Pi_y \eta_x|^2 \right) \nonumber \\[1mm] 
	&+ K_3 \left( \Pi_x\eta_x \right)^{\ast} \left(\Pi_y \eta_y \right) + K_4 \left( \Pi_x\eta_y \right)^{\ast} \left(\Pi_y \eta_x \right)  + c.c.  , \label{eqn:gl-f}	
\end{align}
where $ a(T) = a' (T-T_c) $ and the coefficients $ a'$, $b_{1,2,3}$ and $K_{1,2,3,4} $ are positive and $ T_c $ is the bulk critical temperature. The covariant gradient is defined as
$\bm{\Pi}= -i\hbar \bm{\nabla}  - 2e \bm{A} / c $ with $ \bm{A} $ the vector potential, $ e $ and $c $ the electron charge and the speed of light, respectively (from now on $\hbar = 1$). Furthermore, we assume that due to $ L_z \gg L $ we can ignore any $z$-dependence of the superconducting order parameter. All quantities are considered per unit length along the $z$-direction. The homogeneous bulk state is obtained by minimizing $F$ for $ T < T_c $. With $ \bm{\eta} = \eta_0 (1, \chi i) $, where $ \chi= \pm 1 $ denotes the chirality, the bulk value is given by 
\begin{equation}
|\eta_0|^2 = \frac{-a(T)}{4b_1-b_2+b_3} \; .
\end{equation}
The gradient terms allow us to derive the expressions of the supercurrents in the plane of the film. Using $\bm{j} = - c \partial F / \partial \bm{A}$ we find
\begin{align}
	j_x &= 2e  \Big[ K_1 \eta_x^{\ast} \Pi_x \eta_x + K_2 \eta_y^{\ast} \Pi_x \eta_y \nonumber \\ 
	&\hspace{1cm}+ K_3 \eta_x^{\ast} \Pi_y \eta_y + K_4 \eta_y^{\ast} \Pi_y \eta_x + c.c. \Big], \label{eqn:jx} \\
	j_y &= 2e   \Big[ K_2 \eta_x^{\ast} \Pi_y \eta_x + K_1 \eta_y^{\ast} \Pi_y \eta_y \nonumber \\ 
	&\hspace{1cm}+ K_3 \eta_y^{\ast} \Pi_x \eta_x + K_4 \eta_x^{\ast} \Pi_x \eta_y + c.c. \Big]. \label{eqn:jy}
\end{align}
For $ L \ll \lambda_L $ screening effects are negligible such that we can safely neglect the vector potential from now on.

\subsection{Symmetry dependence of the critical current}
We now turn to the geometry with a finite width in $y$-direction. The two film edges are pair breaking through scattering and yield a local suppression of the order parameter, which is, generally, stronger for $\eta_y$ than $ \eta_x $~\cite{sigrist1991phenomenological}. Considering a current flow along the $x$-direction we introduce a variational ansatz for the order parameter, which takes the essential properties into account:
\begin{align}
	\bm{\eta} = \begin{pmatrix} \tilde{\eta}_x g_x(y) \\[1mm] i \chi \tilde{\eta}_y g_y(y) \end{pmatrix} e^{i\alpha x}. \label{eqn:film_ansatz}
\end{align}
The functions $ g_x (y) $ and $ g_y (y) $ give the transverse spatial dependence for $ -L/2 \leq y \leq +L/2$, and the parameter $ \alpha $ denotes the phase gradient along $x$-direction describing the supercurrent. Moreover, $ \bm{\eta}_0 = (\tilde{\eta}_x , \tilde{\eta}_y ) $ quantifies the magnitude of the two order parameter components (both are positive). The functions $ g_{x,y} $, which satisfy the boundary conditions at the edges due to pair breaking, are converging to 1 deep inside the film assuming that the healing lengths (coherence lengths) are shorter than~$ L $. 
\begin{figure}[t!]
 \centering
	\includegraphics[width=0.75\columnwidth]{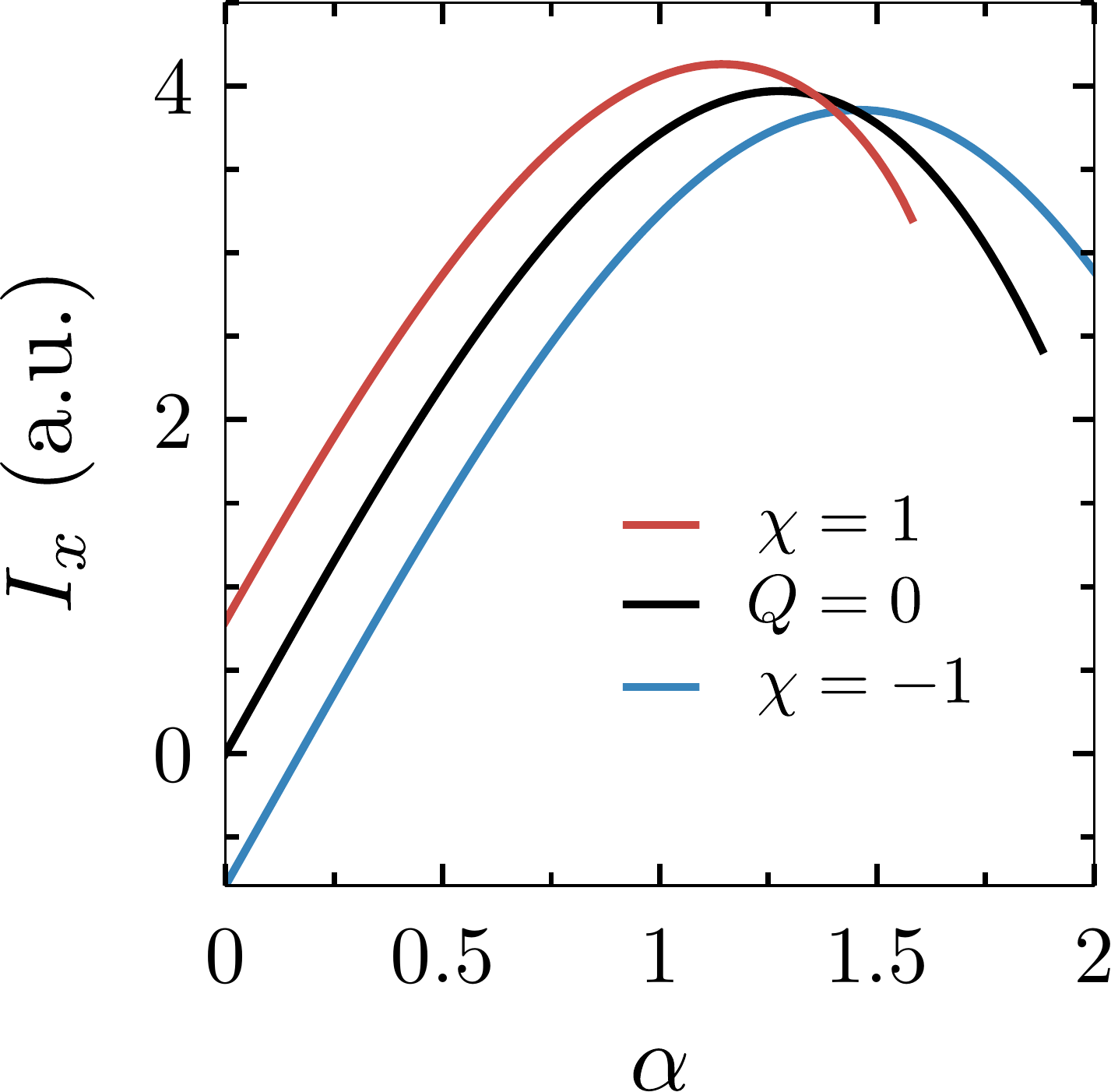}
\caption{The total current in $x$-direction of the thin film as a function of the phase gradient $\alpha$, if the film is symmetric under $\sigma_y$ (black) or not (red and blue).
In the fully symmetric system the solutions for chiralities $\chi=\pm 1$ are degenerate. We used for the Ginzburg-Landau coefficients ($b_1$, $b_2$, $b_3$) = ($0.3 b$, $0.4 b$, $0.2 b$) and ($K_1$, $K_2$, $K_3$, $K_4$) = ($0.6 K$, $0.4 K$, $0.4 K$, $0.4 K$), which are in line with a generic 2d Fermi surface. The integrals [Eq.~(\ref{eqn:integrals})] were taken to be ($A^x$, $A^y$, $B^x$, $B^y$, $B^{xy}$, $C^x$, $C^y$, $D^x$, $D^y$) $\approx$ ($0.95$, $0.6$, $0.7$, $0.5$, $0.36$, $0.01$, $0.05$, $0.4$, $0.0$). Note that our parameters are connected to the order parameter magnitude through $|\eta_0|^2 = -a/b$ and to the coherence length through $\xi^2 = -K/a$.
}
\label{fig:film}
\end{figure}\par
This variational ansatz can now be inserted into the Ginzburg-Landau free energy [Eq.~(\ref{eqn:pwaveExp})] 
and leads to the free energy density per unit length along $x$-direction, 
\begin{widetext}
\begin{gather}
	\frac{f}{L}= \left[ (a + K_1 \alpha^2)A^x + K_2 C^x \right]  \tilde{\eta}_x^2 + \left[  (a + K_2 \alpha^2)A^y + K_1 C^y \right] \tilde{\eta}_y^2  + (2b_1-b_2+b_3) B^{xy} \tilde{\eta}_x^2 \tilde{\eta}_y^2 \nonumber \\[1mm]
	+ b_1 B^x \tilde{\eta}_x^4 + b_1 B^y \tilde{\eta}_y^4 + 2 (K_3D^x-K_4D^y) \alpha \chi \tilde{\eta}_x \tilde{\eta}_y. \label{eqn:film_free}
\end{gather}
\end{widetext}
Here we introduced parameters obtained from the integration over $ y \in  [- L/2, L/2]$,
\begin{gather}
	A^x = \int g_x^2 \frac{dy}{L},  \quad A^y = \int g_y^2 \frac{dy}{L}, \quad B^x = \int g_x^4 \frac{dy}{L}, \nonumber \\[1mm] 
	   B^y = \int g_y^4 \frac{dy}{L}, \quad B^{xy} =  \int g_x^2 g_y^2 \frac{dy}{L}, \quad C^x = \int g_x'^2   \frac{dy}{L}, \nonumber \\[1mm] 
	   C^y = \int g_y'^2   \frac{dy}{L}, \quad  D^x = \int g_x g_y' \frac{dy}{L}, \quad D^y = \int g_y g_x' \frac{dy}{L} .\label{eqn:integrals}
\end{gather} 
 Taking a look at the structure of $f$ [Eq.~\eqref{eqn:film_free}] it is easy to see that the different suppression of the two order parameter components at the edges will generally lead to a double transition, where first $ \eta_x $ and at a lower temperature also $ \eta_y $ nucleate~\cite{sigrist1991doubletransition}. In the following, we assume that the temperature $T$ is sufficiently low such that for $ \alpha =0 $ both order parameters are finite.

Next we derive the current flow for the given variational state. Inserting Eq.~(\ref{eqn:film_ansatz}) into the expressions for the current density [Eqs.~(\ref{eqn:jx},~\ref{eqn:jy})], we find that $ j_y $ vanishes as expected. The total current in $x$-direction takes the form
\begin{gather}
I_x = \int j_x(y) dy = 4e \big[\underbrace{(K_1 \tilde{\eta}_x^2 A^x + K_2 \tilde{\eta}_y^2 A^y) \alpha}_{I_1(\alpha)/4e}  \nonumber \\[1mm] 
 + \underbrace{(K_3D^x - K_4 D^y) \chi \tilde{\eta}_x \tilde{\eta}_y}_{I_2(\alpha)/4e} \big] = I_1(\alpha)+I_2(\alpha).
\label{eqn:currents-film}
\end{gather}
This current is also obtained from $ f $ by a simple derivative with respect to $ \alpha $, $ I_x = 2e  \partial f / \partial \alpha $.

Let us analyze these results by considering the symmetry aspects of our system. If both edges are identical, the film is invariant under the mirror operation $ \sigma_y $. The spatial dependence of the order parameter is then also symmetric under $ y \to - y $: $ g_x(-y)=g_x(y) $ and $ g_y(-y) = g_y (y) $. In this case, the integrals $ D^x $ and $ D^y $ in Eq.~(\ref{eqn:integrals}) are zero, because $ g_x(y) g'_y(y) = -  g_x(-y) g'_y(-y) $ and $ g_y(y) g'_x(y) = -  g_y(-y) g'_x(-y) $. As a consequence, the last term in the free energy density [Eq.~(\ref{eqn:film_free})] and $ I_2 $ [Eq.~(\ref{eqn:currents-film})] vanish. All other integrals are finite and positive.

Considering the behavior of the current as a function of the phase gradient $ \alpha $, we also have to take into account that the order parameter is modified by the current. Minimizing the free energy density with respect to both $ \tilde{\eta}_x $ and $ \tilde{\eta}_y $, we find that a growing $ | \alpha | $ reduces both order parameters. Since we also derive that in this situation $ \tilde{\eta}_{x,y} (\alpha) = \tilde{\eta}_{x,y} (-\alpha) $, we obtain after inserting the order parameters in Eq.~(\ref{eqn:currents-film}) that $ I_1(\alpha) = - I_1(- \alpha) $. The maximum $ I_{c0} = {\rm max}_{\alpha} |I_1(\alpha)| = |I_1(\alpha_0)| $ is hence independent of the current direction and has close to $\alpha_0$ the approximative form $I_x (\alpha)   = I_{c0} + \mathcal{C} (\alpha - \alpha_0)^2 $ with the constant $\mathcal{C} < 0 $. A current larger than $I_{c0}$ includes a normal current and leads to phase slips of the order parameter, which are dissipative and omitted here in our treatment.

If the two edges of the film are different, we find $ g_{x,y}(-y) \neq g_{x,y} (y) $. Consequently, the integrals $ D^{x,y} $ in the last term in $ f/L $ [Eq.~\eqref{eqn:film_free}] and thus also $ I_2 $ [Eq.~\eqref{eqn:currents-film}] are finite. 
Assuming that the term $ Q = (K_3 D^x - K_4 D^y) $ is small, the order parameter components are only slightly corrected for given $ \alpha $, 
\begin{align}
	 \tilde{\eta}_{\mu} (\alpha) \to  \tilde{\eta}_{\mu} (\alpha) + \delta \tilde{\eta}_{\mu},
\end{align}
with $\mu=\{x,y\}$. We derive the correction by expanding the free energy around its maximum at $\alpha_0$ in terms of $ \delta \tilde{\eta}_{\mu}$. The $Q$-term is treated thereby as a small perturbation. In lowest-order correction we find 
\begin{align}
\delta \tilde{\eta}_{\mu} \approx R_{\mu} Q \chi \alpha , 
\end{align}
where the factors $ R_{\mu}  = R_{\mu} (\alpha) $ depend on $ \alpha^2 $.\par 
Inserting this into the expression of the total current [Eq.~(\ref{eqn:currents-film})], which we approximate qualitatively around its unperturbed minimum at $\alpha_0 > 0$ through a parabola, takes us to 
\begin{align}
I_x (\alpha) &\approx  I_{c0} + \mathcal{C}(\alpha - \alpha_0)^2 + 8e \big[ K_1 A^x \tilde{\eta}_x \delta \tilde{\eta}_x \nonumber \\[1mm] 
&\hspace{1cm} + K_2 A^y \tilde{\eta}_y \delta \tilde{\eta}_y \big] \alpha_0 + Q  \tilde{\eta}_x \tilde{\eta}_y \chi \nonumber \\[1mm] 
&= I_{c0} + \mathcal{C}(\alpha - \alpha_0)^2 + K(\alpha_0) Q \chi \alpha + \tilde{Q}(\alpha_0) \chi \nonumber \\[1mm] 
&= I_{c0} + \mathcal{C} (\alpha - \tilde{\alpha}_0)^2 + \delta I_c,
\end{align}
where $\mathcal{C}$ is an undetermined constant and $ \tilde{\alpha}_0 $ the new position of the maximal current. We introduced 
\begin{align}
	K(\alpha_0) = 8e \alpha_0 \big[ K_1 A^x \tilde{\eta}_x(\alpha_0) R_x + K_2 A^y \tilde{\eta}_y(\alpha_0) R_y \big],
\end{align}
and
\begin{align}
	\tilde{Q}(\alpha_0) = Q \tilde{\eta}_x(\alpha_0) \tilde{\eta}_y(\alpha_0).
\end{align}
Through comparison we derive that the maximal current is shifted by 
\begin{align}
	\delta I_c = \left[K(\alpha_0) Q + \tilde{Q}(\alpha_0)\right] \chi,
\end{align}
which depends on the chirality $\chi = \pm 1$. Our result shows that the critical current for a particular direction is influenced by the sign of the chirality. This is equivalent to having to a different critical current for positive and negative directions with fixed chirality,
\begin{equation}
I_c^{+} - I_c^{-} = \delta I_c = \left[K(\alpha_0) Q + \tilde{Q}(\alpha_0)\right] \chi ,
\end{equation}
as we will show below. The relevant conditions for this difference are broken time-reversal symmetry and the lack of mirror symmetry transverse to the film ($ y \to - y $). Note that the observation of $ I_c^+ \neq I_c^- $ could be used to detect the onset of a time-reversal-symmetry-breaking phase. For example, an order parameter, which is given by $ \bm{\eta} = (\eta_x, 0) $ for $ T > T^* $ and changes into $ \bm{\eta} = (\eta_x , \eta_y )$ for $ T < T^* $~\cite{sigrist1991doubletransition}, leads to the described splitting of the critical currents, if the relative phase between the components is $ \chi \pi/2 $.

To complement the approximative treatment we show in Fig.~\ref{fig:film} the result of the numerically determined $ \alpha $-dependence of the total current, which we obtain by the minimization of $ f/L $ [Eq.~\eqref{eqn:film_free}] with respect to the two order parameter components for given $ \alpha$. We see for positive $ \alpha $ a splitting of the two current maxima if $ Q \neq 0 $, which correspond to opposite chirality. Therefore, this splitting represents the non-reciprocal behavior of the critical current. Considering, for instance, the current for $ \chi = -1 $ in Fig.~\ref{fig:film} the time-reversal operation will change the sign of $ \chi $ and of the current. Thus, the maximal current of $ \chi=-1 $ corresponds to $ - I_c^- $ of the reversed current for $ \chi=+1 $. We also observe that compared to $ Q=0 $ a non-vanishing value of $Q$ leads to an offset of the current at $ \alpha = 0 $. The direction of this net current depends on the sign of $ Q \chi $, since we have $ I_x (\alpha =0 ) = I_2 (\alpha =0 ) $ due to $ I_1(\alpha=0) = 0 $. Fig.~\ref{fig:film} also shows a growing difference in the suppression of the current for increasing $ \alpha $ beyond the maximal value, which is connected with the decrease of the order parameter.

\section{Josephson junction between chiral superconductors}

\begin{figure}[t!]
 \centering
	\includegraphics[width=0.99\columnwidth]{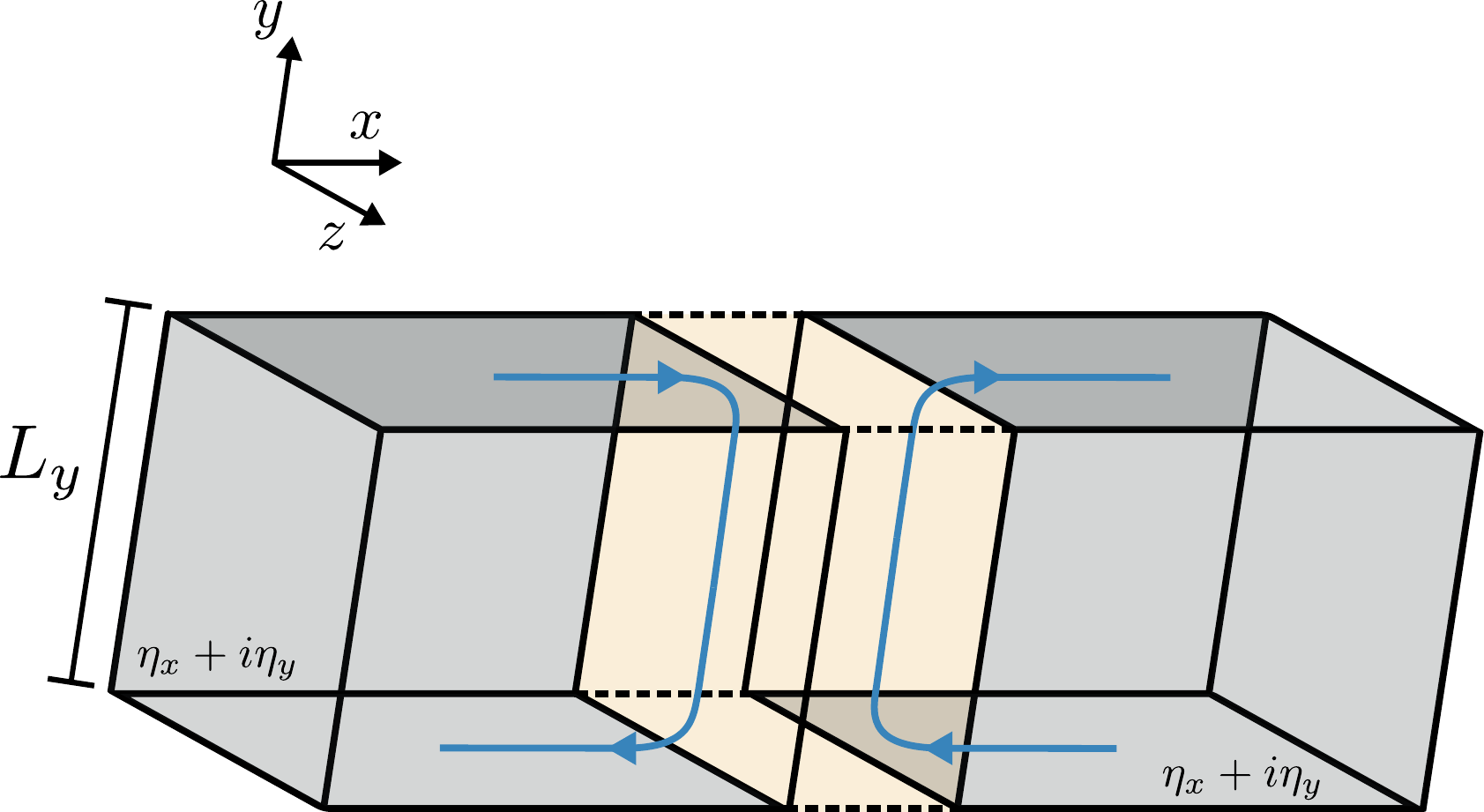}
\caption{Schematic picture of a Josephson junction consisting of two chiral superconductors with the same chirality (grey) and a normal state region (yellow). The Josephson current (blue) flows along the edges. The system is homogeneous in the $z$-direction.
}
\label{fig:JJ_sketch}
\end{figure}

In this section, we study the conditions for non-reciprocal critical currents in a Josephson junction connecting two identical, chiral superconductors. For a spatially extended junction we derive the effective free energy functional of the Josephson phase difference $ \phi $. Compared to the standard theory chiral order parameters yield an additional term in the free energy, which describes spontaneous currents at the junction edges, influences the line energy of Josephson vortices and splits the critical currents running in opposite directions in the absence of a specific mirror symmetry.

\subsection{Effective free energy functional of the junction}

We consider a Josephson junction positioned at $ x=0 $ that extends in $y$- and $z$-direction, as Fig.~\ref{fig:JJ_sketch} illustrates. The coupling between the two superconductors is implemented phenomenologically by combining the order parameters from both the left ($l$) and right ($r$) sides, 
\begin{equation}
\bm{\eta} (\bm{r}) = \bm{\eta}^l (\bm{r}) +  \bm{\eta}^r (\bm{r}),
\end{equation}
which overlap in the junction region. Analogously to the thin film we introduce a variational form for the spatial dependence of the order parameter components,
\begin{align}
    \bm{\eta}^l(\bm{r}) &= \eta_0 \begin{pmatrix} g_x^l(\bm{r}) \\ i\chi_l g_y^l(\bm{r}) \end{pmatrix} e^{i\phi(y,z)/2}, \label{eqn:etal}\\[1mm]
    \bm{\eta}^r(\bm{r}) &= \eta_0 \begin{pmatrix} g_x^r(\bm{r}) \\ i\chi_r g_y^r(\bm{r}) \end{pmatrix} e^{-i\phi(y,z)/2}, \label{eqn:etar}
\end{align}
with non-negative $ g_x^j(\bm{r}) \neq g_y^j(\bm{r}) $ ($j=\{l,r\}$), the phase $ \phi = \phi^l - \phi^r $ and the chirality $ \chi_{l,r} = \pm 1 $. We assume the following symmetry property,
\begin{equation}
g_{\mu}^l(x,y,z) = g_{\mu}^r(-x,y,z) =  \begin{cases} 0 & x \to + \infty \\[1mm] 1 & x \to - \infty \end{cases}, \label{eqn:g-sym}
\end{equation}
for both $\mu = x$ and $y$. The overlap of the interpenetrating order parameters is restricted to a narrow region around $ x =0 $ of the order of the coherence length. This spatial dependence implies invariance of $ g_{\mu}^l(x,y,z) + g_{\mu}^r(x,y,z) $ for the mirror operation $ \sigma_x $, taking the junction as the mirror plane.

With this setup we can now derive the free energy functional for $ \phi $ using the Ginzburg-Landau theory as given in Eq.~(\ref{eqn:pwaveExp}). 
Details on the derivations can be found in Appendix~\ref{app:A}. For simplicity, we neglect any $z$-dependence and assume a sufficiently long extension $ L_z $. Similar to the case of the thin film we can avoid demagnetization fields and consider only magnetic fields parallel to the $z$-axis. The free energy functional is then given by
\begin{align}
	F_{J}[\phi] =  &\frac{\Phi_0^2 L_z }{16 \pi^3 w_J}  \int_{-L_y/2}^{+L_y/2} dy  \Bigg[ \frac{1}{2}\left( \frac{\partial \phi}{\partial y} \right)^2  + \lambda_J^{-2} (1- \cos\phi) \nonumber \\
	&+ \gamma(\chi_l, \chi_r) \lambda_J^{-2}  \frac{\partial \phi}{\partial y} (1-\delta_{\gamma}\cos\phi) \Bigg],\label{eqn:energy_J}
\end{align}
where $ \Phi_0 $ denotes the flux quantum and $ \lambda_J $ the Josephson penetration depth,
\begin{equation}
\lambda_J^{-2} = \frac{8 \pi^2 w_J j_c}{\Phi_0 c},
\end{equation}
which depends on the critical current density of the junction $ j_c $.

The derivative of the phase is related to the local ``magnetic field'' (magnetic flux density) in the junction through 
\begin{equation} 
\frac{\partial \phi}{\partial y} = \frac{2 \pi}{\Phi_0} B_z (y) w_J ,
\end{equation} 
where $ w_J \approx d_J + 2 \lambda_L $ corresponds to the effective magnetic width and $ d_J $ to the microscopic width of the junction. The first term of Eq.~\eqref{eqn:energy_J} represents the magnetic field energy, while the usual coupling between the two superconductors is described by junction energy in the second term. We ignore higher order couplings as they are not essential for our discussion. Note that these first two terms of Eq.~\eqref{eqn:energy_J} are standard for Josephson junctions and that their derivation can be found in textbooks such as Ref.~\cite{barone1982physics}.\par 
However, ``non-standard'' is the third term in Eq.~(\ref{eqn:energy_J}), which describes the anomalous behavior of a Josephson junction between two chiral superconductors. We will call it the $ \gamma$-term from now on. Using our variational order parameters we derive this contribution to the functional through the gradient terms in the Ginzburg-Landau functional [Eq.~(\ref{eqn:pwaveExp})]. This leads to the relation, 
\begin{equation}
\gamma(\chi_l, \chi_r) \lambda_J^{-2} \frac{\Phi_0^2 }{16 \pi^3 w_J} = -\eta_0^2 A_{\gamma} (\chi_l + \chi_r),  \label{eqn:gam}
\end{equation}
with 
\begin{equation}
A_{\gamma} = \int_{-\infty}^{+\infty} dx \; \left[ K_3 g_x^l (x)' g_y^l(x) - K_4  g_y^l (x)' g_x^l(x) \right], \label{eqn:Ag}
\end{equation}
and $ \delta_{\gamma} = -B_{\gamma} / A_{\gamma} $ with
\begin{equation}
B_{\gamma} = \int_{-\infty}^{+\infty} dx \; \left[ K_3 g_x^l (x)' g_y^r(x) - K_4  g_y^l (x)' g_x^r(x) \right] , \label{eqn:Bg}
\end{equation}
where we ignore any $y$-dependence of $g_{\mu}^{l,r} (\bm{r})$ for simplicity. 
It is obvious that the $\gamma$-term behaves like an ``external field'' $ \tilde{H}_z(\phi) $ coupling to the local magnetic field $ B_z(y) $,
\begin{equation}
 \tilde{H}_z(\phi) = -\frac{\gamma \Phi_0}{2 \pi \lambda_J^2 w_J} (1- \delta_{\gamma} \cos \phi) ,  \label{eqn:hz}
 \end{equation}
such that  the $\gamma$-term reduces to 
\begin{equation}
 F_{\gamma} = \int dy \frac{\tilde{H}_z(\phi) B_z w_J}{4 \pi} .
\end{equation}
The physical origin of this field can be attributed to the spontaneous current density $ j_y (x) $ running on the two sides of the Josephson junction in opposite $ y $-directions, if the two chiralities $ \chi_{l,r} $ are identical, i.e. $ j_y(x) = - j_y (-x) $ (see also Fig.~\ref{fig:JJ_sketch} and Appendix~\ref{app:A}). For $ \chi_{l} = - \chi_r $ the current density is symmetric, $ j_y(x) =  j_y (-x) $. The generated magnetic field distribution is then odd with respect to $x$ and cancels to zero such that $ \gamma = 0 $. From a symmetry point of view, the operation $ {\cal T} \sigma_x $ leaves both the current-induced $ \tilde{H}_z( \phi) $ and the superconducting states invariant for $ \chi_{l} \neq \chi_r $.  Obviously, this property is reflected in the simple $ \chi_{l,r} $-dependence of $ \gamma $ in Eq.~(\ref{eqn:gam}).

Another important feature of the $\gamma$-term is that it corresponds to a total differential,
\begin{equation}
 \frac{\partial \phi}{\partial y} (1- \cos \phi ) = \frac{\partial}{\partial y} ( \phi - \sin \phi ),
\end{equation}
and, if $ \gamma $ and $ \lambda_J $ are constant, does not enter the variational equation for $\phi$, but only the boundary condition. The variatonal equation of $ F_J $ has the usual form, 
\begin{align}
	\frac{\partial^2 \phi}{\partial y^2} = \frac{1}{\lambda_J^2} \sin\phi, \label{eqn:var_bulk}
\end{align}
while the boundary conditions are modified,
\begin{align}
	 &\left[ \frac{\partial \phi}{\partial y}  +\frac{\gamma}{\lambda_J^2}  \left(1-\delta_{\gamma}\cos\phi \right) \right] \Bigg |_{y=-\frac{L_y}{2}} \nonumber \\[1mm] 
	 &\hspace{3.5cm}= \frac{2\pi w_J}{\Phi_0} H_z\left(-\frac{L_y}{2}\right), \label{eqn:var_bound1} \\[2mm]
	 &\left[ \frac{\partial \phi}{\partial y}  +\frac{\gamma}{\lambda_J^2}  \left(1-\delta_{\gamma}\cos\phi \right) \right] \Bigg |_{y=+\frac{L_y}{2}} \nonumber \\[1mm] 
	 &\hspace{3.5cm}= \frac{2\pi w_J}{\Phi_0} H_z\left(+\frac{L_y}{2} \right), \label{eqn:var_bound2}
\end{align}
where we introduce the magnetic fields $ H_z (\pm L_y / 2 ) $ at the boundaries of the junction ($ y = \pm L_y/2 $). This magnetic field can be decomposed into
\begin{equation}
H_z \left( \pm \frac{L_y}{2} \right) = H_e \mp \frac{2 \pi}{c} \frac{I}{L_z} ,
\end{equation}
where $ H_e $ denotes the net external magnetic field and $ I $ the net current through the junction. Prior to our discussion of the critical current we consider in the following first a homogeneous junction without any net external magnetic field and current, i.e. $ H_e = I = 0$. Our aim is to illustrate the impact of the $\gamma$-term on the energy of a Josephson vortex and the presence of spontaneous currents at the ends of the junction.

\subsection{Josephson vortices}

A Josephson vortex corresponds to a kink (anti-kink) soliton-like solution of the differential equation [Eq.~(\ref{eqn:var_bulk})] given by
\begin{align}
	\phi_{\pm} (y) = 4 \tan^{-1} \left( e^{\pm\frac{y-y_0}{\lambda_J}} \right),
\end{align}
as shown for example in textbooks such as Ref.~\cite{barone1982physics}. This vortex, which is centered at position $ y_0 $ and includes a magnetic flux $ \pm \Phi_0 $, extends along (perpendicular to) the junction over a length of $ \lambda_J $ ($ w_J $). We now insert this kink solution into the free energy functional and integrate over $ y $ assuming $ L_y \gg \lambda_J $. The first two terms give the usual energy of a Josephson vortex per unit length along the $z$-direction,
\begin{equation}
F_v = \frac{ \Phi_0^2}{2 \pi^3 w_J}\frac{1}{\lambda_J}, 
\end{equation}
which is independent of the orientation of the vortex (kink or anti-kink). The correction due to the $\gamma$-term is then straightforwardly calculated:
\begin{align}
\Delta F_{v, \pm}  & = \frac{ \Phi_0^2}{16 \pi^3 w_J}  \int dy \;  \frac{\gamma}{\lambda_J^2} \frac{\partial \phi_{\pm}}{\partial y} (1 - \delta_\gamma  \cos \phi_{\pm}), \nonumber \\[2mm] & = 
\frac{ \Phi_0^2}{16 \pi^3 w_J}   \int dy \; \frac{\gamma}{\lambda_J^2} \frac{\partial }{\partial y} ( \phi_{\pm} - \delta_{\gamma} \sin \phi_{\pm} ) ,\nonumber \\[2mm]
& = \frac{ \Phi_0^2}{16 \pi^3 w_J} \frac{\gamma}{\lambda_J^2} ( \phi_{\pm} - \delta_{\gamma} \sin \phi_{\pm} ) \Big |_{y= -\infty}^{y= + \infty}, \nonumber  \\[2mm]
&= \pm \frac{ \Phi_0^2}{8 \pi^2 w_J} \frac{\gamma}{\lambda_J^2}.
\end{align}
This shows that there appears a difference in vortex line energy depending on the orientation of the vortex, parallel or antiparallel to the chirality. Thus, through the $\gamma$-term vortices may be more easily created or stabilized than anti-vortices or vice versa depending on $ \chi=\chi_r = \chi_l $.

\subsection{Spontaneous edge currents}
\begin{figure}[t!]
 \centering
	\includegraphics[width=0.65\columnwidth]{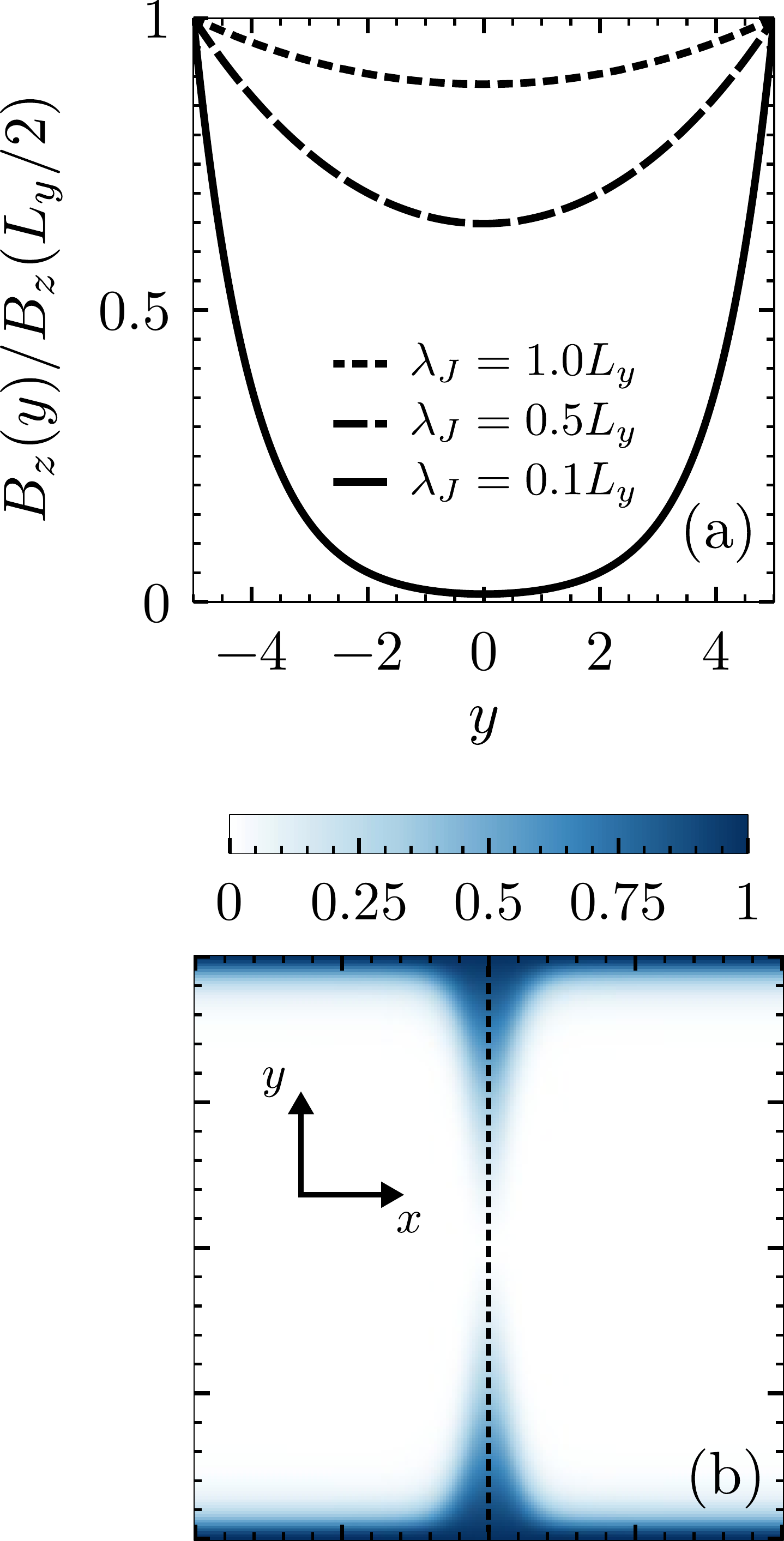}
\caption{(a) The normalized magnetic flux density $B_z(y) \propto \partial \phi / \partial y$ created by spontaneous edge currents for $H_e = 0$ and $\gamma \ll 1$. Latter is assumed to be small enough such that $\phi \ll 1$. Taking $L_y = 10$ we consider different values of $\lambda_J$. (b) Schematic illustration of the spontaneous magnetic field distribution inside of the Josephson junction. The field values are normalized and the location of the junction is given by the dashed line. 
}
\label{fig:spont_phideriv}
\end{figure} 
Here we address a second feature due to the $ \gamma $-term that appears in a junction of finite extension $ L_y $. For simplicity we assume that $ \gamma \ll 1 $. The deviation from the usual lowest energy solution, $ \phi = 0 $ (or an integer multiple of $ 2 \pi $), is then very small, i.e. $ | \phi | \ll 1 $. Thus, the variational equation [Eq.~(\ref{eqn:var_bulk})] can be approximated as
\begin{equation}
\frac{\partial^2 \phi}{\partial y^2} = \frac{\phi}{\lambda_J^2} ,
\end{equation}
which is solved by 
\begin{equation}
\phi(y) = C_+ e^{+y / \lambda_J} + C_- e^{-y / \lambda_J} .
\end{equation}
The boundary conditions ($ H_e = I = 0 $) combined with the approximation $ \cos \phi \approx 1 $ yield the solution
\begin{equation}
\phi(y) = - \gamma (1 - \delta_{\gamma} ) \frac{\sinh (y / \lambda_J)}{\cosh (L_y/2 \lambda_J)} .
\end{equation}
This leads to a spontaneous Josephson current,
\begin{equation}
j(y) = j_c \sin \phi(y) \approx - \gamma j_c (1 - \delta_{\gamma} ) \frac{\sinh (y / \lambda_J)}{\cosh (L_y/2 \lambda_J)},
\end{equation}
corresponding to a current pattern which originates from both edges of the junctions, extends over a length $ \lambda_J $ and runs in opposite directions on the two edges. This is analogous to the edge currents at ordinary surfaces that are confined within the length of the London penetration depth $ \lambda_L \ll \lambda_J $. 
These currents give rise to a magnetic flux penetrating at the edges,
\begin{equation}
B_z (y) = - \frac{\Phi_0\gamma}{2\pi w_J \lambda_J}  (1 - \delta_{\gamma})  \frac{\cosh (y / \lambda_J)}{\cosh (L_y/2 \lambda_J)} ,
\end{equation}
as shown in Fig.~\ref{fig:spont_phideriv}~(a). An extended view of the junction and its environment is depicted schematically in Fig.~\ref{fig:spont_phideriv}~(b), where we observe that the well-confined spontaneous magnetic flux extends further into the superconductor along the junction.

\subsection{Non-reciprocal critical currents}
We now turn to the situation where the lack of a specific mirror symmetry in this type of Josephson junction causes a difference between the critical currents flowing in opposite directions. We consider a setup where the junction is invariant under $ {\cal T} \sigma_x  $, but not under $ {\cal T} \sigma_y $. This may be implemented, for instance, by assuming different Josephson couplings for positive and negative $ y $, as in Eq.~(\ref{eqn:split-junction}). We first take a look at a short junction to show how the non-reciprocal critical current arises within a simple approximative, yet analytical treatment. Afterwards, we will solve numerically the case of an extended junction, where the supercurrent is limited by the entry of Josephson vortices from the boundary into the junction. 

\subsubsection{Short-junction limit}

In the limit $ L_y \ll \lambda_J $, the junction is considered as ``short'' and the variational equation for the phase is approximately given by
\begin{equation}
\frac{\partial^2 \phi}{\partial y^2} = 0 ,
\end{equation}
which is solved by the ansatz $ \phi(y) = \alpha + by $. We now assume that 
\begin{equation}
\lambda_J^{-2}(y) = \begin{cases} \lambda_+^{-2} & -\frac{L_y}{2} \leq y < 0 \\[2mm]
\lambda_-^{-2} & 0 \leq y  \leq \frac{L_y}{2} \end{cases}, \label{eqn:split-junction}
\end{equation}
while we keep $ \gamma $ constant. We then introduce $ q_{\pm} = \gamma \lambda_{\pm}^{-2} = \bar{q} \pm \tilde{q} >0 $. Furthermore, we assume that both $ \bar{q} $ and $ \tilde{q} $ are small ($|\tilde{q}L_y| \ll |\bar{q}L_y| \ll 1$) such that it is justified to use the above simple solution for the whole range of the junction. The boundary conditions [Eqs.~(\ref{eqn:var_bound1},~\ref{eqn:var_bound2})] are rewritten as
\begin{align}
b+ q_+ [1 - \delta_{\gamma} \cos( \alpha + bL_y/2) ] & = - C I, \\[2mm]
b+ q_- [1 - \delta_{\gamma} \cos(  \alpha - bL_y/2) ] & =  C I,
\end{align}
where we set the external magnetic field $ H_e $ to zero and introduce $ C = 4 \pi^2 w_J / \Phi_0 c L_z $. Adding these two equations we obtain
\begin{align}
b & = - \bar{q} [1 - \delta_{\gamma} \cos  \alpha \cos(bL_y/2) ] + \delta_{\gamma} \tilde{q} \sin  \alpha \sin(bL_y/2), \nonumber \\[2mm]
& \approx -  \bar{q} \left(1 - \delta_{\gamma} \cos  \alpha \right) + \frac{b L_y}{2} \delta_{\gamma} \tilde{q} \sin  \alpha, \nonumber \\[2mm]
& \approx  - \bar{q} \left(1 - \delta_{\gamma} \cos  \alpha \right) + \mathcal{O}(\bar{q} L_y \tilde{q}), \label{eqn:b-q}
\end{align}
which is compatible with $ |bL_y| \sim |\bar{q} L_y| \ll 1 $.

This result can now be used to get an expression for the total current as a function of $  \alpha $ by subtracting the two boundary condition,
\begin{align}
C I & =  - \tilde{q}  - \frac{\delta_{\gamma}}{2} [ q_- \cos(\alpha - bL_y/2) - q_+ \cos(\alpha + bL_y/2)], \nonumber \\[2mm]
& \approx -  \tilde{q} \left( 1 - \delta_{\gamma} \cos \alpha \right) - \delta_{\gamma} \bar{q} \frac{b L_y}{2} \sin \alpha , \nonumber \\[2mm]
& \approx  \left( 1 - \delta_{\gamma} \cos \alpha \right) \left( \delta_{\gamma} \bar{q}^2 \frac{L_y}{2} \sin \alpha  -  \tilde{q} \right).
\end{align}
Since for a homogeneous junction the relation $ \tilde{q} = 0 $ holds, we find $ I(-\alpha) = - I(\alpha) $, which leads to the same critical current for positive and negative flow direction. However, for $ \tilde{q} \neq 0 $ the maximal currents for the two flow directions are different, $ I_c^+ \neq I_c^- $. In lowest order we obtain $ I_c^+ - I_c^- \propto \tilde{q} $. Note that this behavior is analogous to the one derived for the thin film in Sect.~\ref{sec:film}.

\subsubsection{Extended junction} 
We now consider the opposite limit $ L_y \gg \lambda_J $, the extended junction, for which it is rather difficult to address analytically the question of an inhomogeneous junction, even in an approximative way. However, the differential equation for homogeneous junction [Eq.~(\ref{eqn:var_bulk})] can be solved using Jacobian elliptic functions,  ${\rm sn}(u | k^2)$, ${\rm cn}(u | k^2)$ and ${\rm dn}(u | k^2)$, which depend on the argument $u$ and parameter $k$ \cite{owenscalapino1967}. Taking $ k \leq 1 $ we express the phase $ \phi $ through
\begin{align}
	\sin \left( \frac{\phi}{2} \right) &= {\rm cn} \left( \frac{y-y_0}{k \lambda_J} \middle| k^2 \right), \\[1mm]
	\cos \left( \frac{\phi}{2} \right) &= -{\rm sn} \left( \frac{y-y_0}{k \lambda_J} \middle| k^2 \right),
\end{align}
whereby the values of $ k $ and $ y_0 $ have to be determined by the boundary conditions. These are formulated as
\begin{gather} 
\frac{1}{k \lambda_J } {\rm dn} \left( \frac{\pm \frac{L_y}{2} -y_0}{k \lambda_J} \middle| k^2 \right) + \frac{\gamma}{\lambda_J^2} 
{\rm cn}^2 \left( \frac{\pm \frac{L_y}{2} -y_0}{k \lambda_J} \middle| k^2 \right) \nonumber \\[1mm]  =  \mp \frac{C}{2} I . \label{eqn:Jac-BS}
\end{gather}
where we again neglect the external magnetic field and assume for simplicity $\delta_{\gamma}=1$. 
Note that there is an analogous set of equations for $ k>1 $. 

Instead of discussing the inhomogeneous junction, we focus now on the homogeneous situation using the solutions given above. We search for the values of $ y_0 $ and $k$ that fulfill the boundary conditions [Eq.~(\ref{eqn:Jac-BS})] and yield the maximal possible current $I$.\par
\begin{figure}[t!]
 \centering
	\includegraphics[width=\columnwidth]{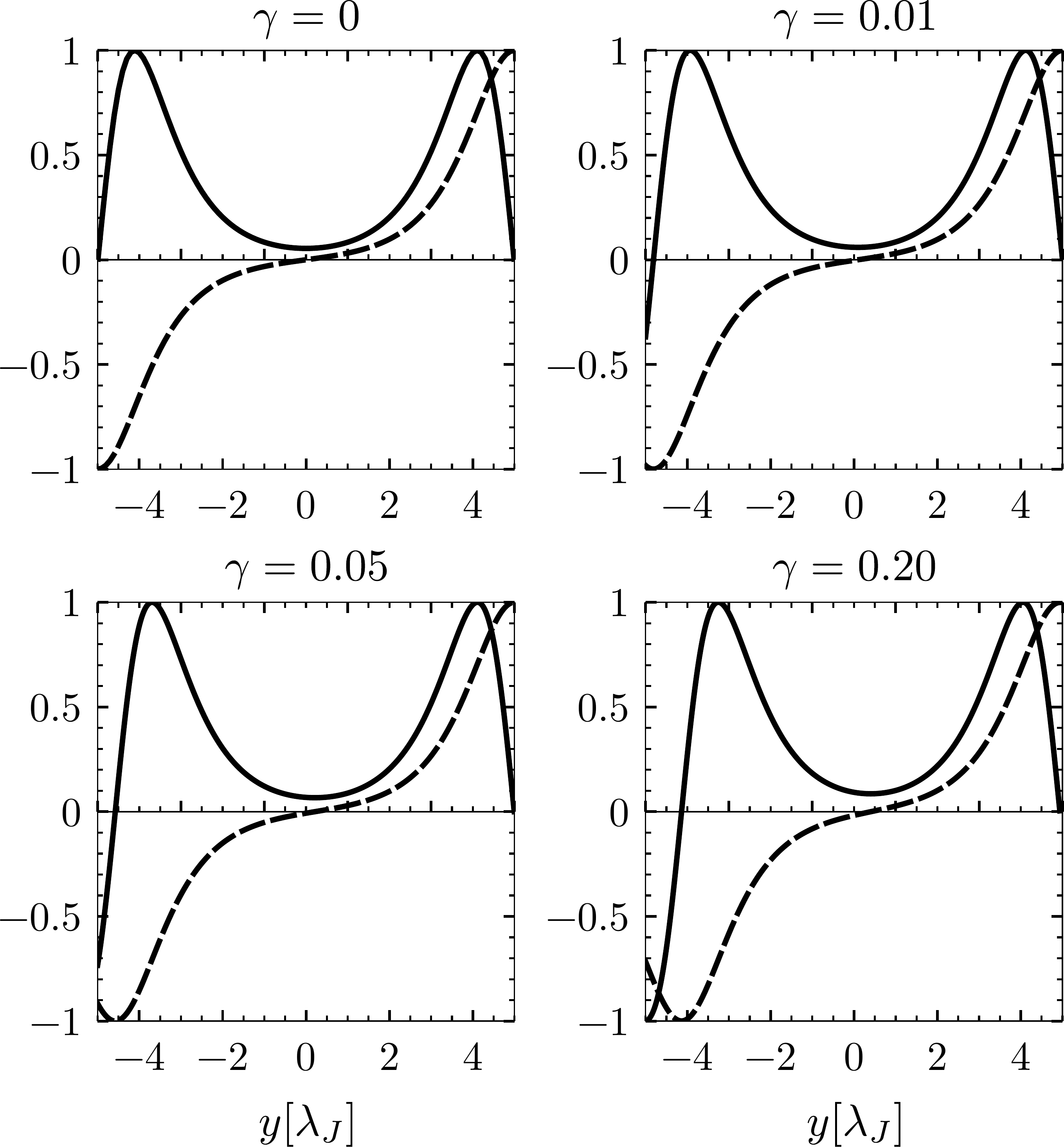}
\caption{The current density (solid) and magnetic field distribution (dashed) inside the junction for $H_e = 0$ and various $\gamma$ values. The distance is expressed in units of the Josephson penetration depth $\lambda_J$. Note that our results converge to the ones contained in \cite{owenscalapino1967} for $\gamma \rightarrow 0$.
}
\label{fig:cur&mag_longJJ}
\end{figure}

  
This allows us to analyze numerically the conditions for the critical current, which is reached when the first Josephson vortex enters the junction. Such vortices then move along the junction, driven by the current, and lead to a finite voltage. The standard case for $ \gamma = 0 $ has been discussed by Owen and Scalapino~\cite{owenscalapino1967} and is included in Fig.~\ref{fig:cur&mag_longJJ}, where we plot the current density and magnetic field distribution in the junction with the possible maximal current (critical current). In this case Josephson vortices nucleate at both ends of the junction under the same conditions. Actually, the situation displayed in Fig.~\ref{fig:cur&mag_longJJ} yields a vortex entering at $y=L_y/2$ and an anti-vortex at $y=-L_y/2$. In the resistive state with a current slightly above the critical current they enter the junction and move towards its center, where they then annihilate each other. Note that it does not matter in which direction the current flows, vortex and anti-vortex would just be exchanged. 

\begin{figure}[h!]
 \centering
	\includegraphics[width=0.48\columnwidth]{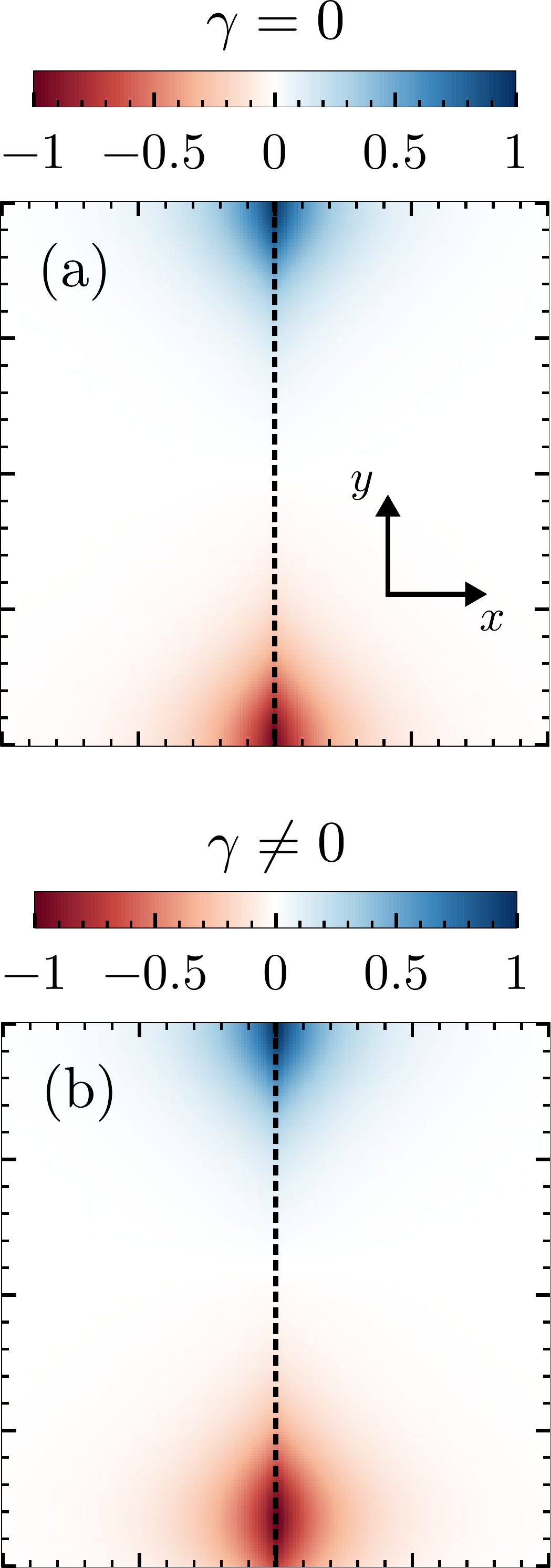}
\caption{The schematic magnetic field pattern if a total current is running through the Josephson junction for (a) $\gamma=0$ and (b) $\gamma \neq 0$. Again we normalized the values by the maximum of the magnetic field. Note that the spontaneous field distribution depicted by Fig.~\ref{fig:spont_phideriv} is negligibly small compared to the magnetic fields plotted here. The position of the junction is indicated by the dashed line. 
}
\label{fig:field_sketch2}
\end{figure}

Turning on the $ \gamma $-term and again evaluating numerically the current density and magnetic field distribution for the maximal supercurrent through the junction, we find the remaining sequence of panels depicted in Fig.~\ref{fig:cur&mag_longJJ} for a current in positive direction. Increasing the values of $ \gamma $ yields a pattern with growing asymmetry between the left and right hand side. It is evident that the nucleation of the first, penetrating vortex occurs on the left side ($y=-L_y/2$). For illustration purposes, the different behavior at both ends of the junction is also depicted qualitatively in Fig.~\ref{fig:field_sketch2}.\par
The origin of this asymmetry is that a Josephson anti-vortex is energetically cheaper than a vortex in the specific situation considered here. Thus, the current is limited by the penetration of anti-vortices from the left-hand side. Reversing the current direction leads to the penetration of the corresponding anti-vortex from the other side.
For a homogeneous junction, i.e. a junction which is symmetric under $ {\cal T} \sigma_y $, the magnitude of the critical current remains the same for both current directions. Naturally, the critical current density is connected with the critical current density $j_c$.

This is now the key to understanding the splitting of $ I_c^+ $ and $ I_c^-$, if $ {\cal T} \sigma_y $ is not a symmetry anymore. We take again our simple model [Eq.~(\ref{eqn:split-junction})], where the junction has a different critical current density for $ y <0 $ and $ y > 0 $. 
In this case, the anti-vortices can more easily enter from the side where $ j_c $ is lower. On the other hand, the direction of the current decides about the orientation of the vortex on the two edges. In this way, we obtain the non-reciprocal behavior, $ I_c^+ \neq I_c^- $, also for the extended junction, because once the vortex (anti-vortex) is inside the junction for currents above the critical current it will move and therefore yield dissipation.

\section{Discussion and Conclusion} \label{sec:conclusion}
Our two examples provide insight into the symmetry conditions and mechanism that give rise to non-reciprocal behavior of the critical current, the so-called spontaneous superconducting diode effect, in a chiral superconductor. The important non-trivial ingredient is broken time-reversal symmetry. Additionally, a specific asymmetry of the sample is required, which in most cases is simply given by the device geometry and the natural inhomogeneity of materials.

In our study, we cover two related mechanisms to limit the supercurrent. One is based on the dissipation through phase slips in thin films and short Josephson junctions. The other is due to the current-driven motion of flux lines (flux flow resistance) nucleating at the device boundary, as we discussed for long Josephson junctions. Reciprocal critical currents rely on the symmetry, which ensures that both current directions experience the same conditions. It is already violated, if there is a spontaneous current density whose distribution is asymmetric such that a net current flow exists, which adds to the external current. Then, the external currents in opposite directions create different conditions to limit
the maximal possible supercurrent. The magnitude of the splitting of $ I_c^+ $ and $ I_c^- $ depends on various circumstances, like the nucleation of phase slips or the ease with which vortices penetrate a sample due to varying surface roughness or curvature. 

Coming back to the example of the 3-K-phase in Sr$_2$RuO$_4$-Ru mentioned in the introduction \cite{hooper2004split}, we can now state that the observation of a non-reciprocal effect in the critical current likely originates from a transition to a time-reversal-symmetry-breaking phase at $ T^* \approx 2.3 $~K \cite{kaneyasu20103K}. The 3-K-phase corresponds to an inhomogeneous superconducting state where the order parameter presumably nucleates at the interface between Sr$_2$RuO$_4$ and Ru-metal inclusions at $ T' \approx 3 $~K. At first, a transition to a time-reversal symmetric phase occurs. Only at $ T=T^* $ we see the onset of an additional order parameter component, which yields the violation of time-reversal symmetry. The superconducting condensate of neighbouring inclusions overlap and form weak links like Josephson junctions leading to a strong drop of resistivity, which may, however, not be completely zero in the absence of percolation of superconductivity in the sample. Nevertheless, in current-voltage characteristics it is possible to observe critical currents corresponding to the ``interruption'' of superconducting paths, if the current through such weak links exceeds a threshold. In the time-reversal-symmetry-breaking phase these weak links most likely satisfy the symmetry conditions we have discussed above and naturally yield a non-reciprocal behavior. Moreover, the whole Josephson network created by connecting the superconducting islands around each Ru-inclusion provides the conditions for a non-reciprocal critical current, as discussed in Ref.~\cite{kaneyasu20103K}. Our discussion suggests that the observation of a ``spontaneous'' non-reciprocal critical current serves as a good indication for a superconducting phase that breaks time-reversal symmetry. We have analyzed the situation for a chiral superconducting phase, because it yields this effect under least stringent conditions. The superconducting diode effect in non-centrosymmetric superconductors is generally not spontaneous, since it requires the application of a magnetic field to break time-reversal symmetry~\cite{ando2020diode, daido2021intrinsic, yuan2021supercurrent}.

We should also mention that non-chiral superconducting order parameters, which break time-reversal symmetry, can give rise to the same phenomenon, if the device setup provides the same conditions as discussed in the introduction in Eqs.~(\ref{T-psi}) and (\ref{sigma-psi}), where $x$ is the direction of the critical current and $y$ perpendicular to it. Within a tetragonal crystal the pairing state with $ s + i d_{xy} $-wave symmetry ($ \psi_{\chi} = \eta_s  + i \chi \eta_d k_x k_y $) satisfies such conditions. This state is composed by two components belonging to different, one-dimensional representations, which generally do not have the same critical temperature. Nevertheless, if both components are finite, they satisfy Eqs.~(\ref{T-psi}) and (\ref{sigma-psi}) with $ \chi =\pm 1 $, which corresponds here not to chirality. Looking at the setups we discussed, the thin film and the Josephson junction, we find that the phenomenology is identical. For another composite state like $ s + i d_{x^2-y^2} $ we would have to rotate the $ x$- and $ y$-direction by $45^{\circ} $ to obtain the same conditions. 
This means that the observation of a spontaneous, non-reciprocal critical current is not restricted to chiral states, but under certain conditions also possible for other, time-reversal-symmetry-breaking superconducting phases.

\section*{Acknowledgements}
We would like to thank M. H. Fischer, Y. Maeno, Y. Liu and X. Q. Mao for helpful discussions. This work was financially supported by the Swiss National Science Foundation (SNSF) through Division II (No. 184739).


\appendix
\section{Derivation of the anomalous Josephson junction energy term}\label{app:A}
The anomalous $ \gamma$-term in Eq.~(\ref{eqn:energy_J}) is obtained from the gradient terms in the Ginzburg-Landau free energy expansion [Eqs.~(\ref{eqn:pwaveExp}) and (\ref{eqn:gl-f})]. In particular, we need to consider the term
\begin{gather}
f_{34} =  K_3 \left( \Pi_x\eta_x \right)^{\ast} \left(\Pi_y \eta_y \right) + K_4 \left( \Pi_x\eta_y \right)^{\ast} \left(\Pi_y \eta_x \right) \nonumber \\
 + c.c.,
\end{gather}
where we insert the order parameter $ \bm{\eta} (\bm{r}) = \bm{\eta}^l (\bm{r}) + \bm{\eta}^r (\bm{r})  $ with the simplified spatial dependence proposed in Eqs.~(\ref{eqn:etal}) and (\ref{eqn:etar}).  After integrating over $x$ we find
\begin{align}
F' & = \int dx \; dy \; f_{34} ,\nonumber \\[1mm]
& = -(\chi_l+\chi_r) \eta_0^2 \int dy \frac{\partial \phi}{\partial y} \left( A_{\gamma}  - B_{\gamma} \cos \phi \right),
\end{align}
where $ A_{\gamma} $ and $ B_{\gamma} $ are given by Eqs.~(\ref{eqn:Ag} ) and (\ref{eqn:Bg}), respectively. Inserting the same form of the order parameter $ \bm{\eta}(\bm{r}) $ into the expression for the current density $ j_y $ yields
\begin{align}
j_y(x) & = \chi_l [ j_{ll}(x) + j_{lr}(x) \cos \phi ] \nonumber \\[1mm] 
& \hskip 1 cm + \chi_r [j_{rr} (x) + j_{rl} (x) \cos \phi ] ,
\end{align}
with 
\begin{equation}
j_{ab}(x) = - 4e \eta_0^2 \left[ K_3 g_y^a(x) {g_x^b}(x)' - K_4 g_x^b (x) {g_y^a}(x)' \right]
\end{equation}
and $ a,b \in \{l,r\} $.

Taking into account the symmetry we imposed in Eq.~(\ref{eqn:g-sym}), $ g_{\mu}^l(x) = g_{\mu}^r(-x) $, where we restrict ourselves to the $x$-dependence, we find that $ j_{ll}(x) = - j_{rr}(-x) $ and $ j_{lr}(x) = - j_{rl} (-x) $. The current distribution is even for
\begin{gather}
j_y^{({\rm even})}(x)  = \frac{1}{2} [ j_y(x) + j_y(-x) ],  \\[1mm]
 = \frac{\chi_l - \chi_r}{2} \left\{ j_{ll}(x) - j_{rr} (x) + \left[j_{lr}(x) - j_{rl}(x) \right] \cos \phi \right\}, \nonumber 
\end{gather}
and odd for 
\begin{gather}
j_y^{({\rm odd})}(x)  = \frac{1}{2} [ j_y(x) - j_y(-x) ],  \\[1mm]
 = \frac{\chi_l + \chi_r}{2}  \left\{ j_{ll}(x) + j_{rr} (x) + \left[j_{lr}(x) + j_{rl}(x) \right] \cos \phi \right\}. \nonumber 
\end{gather}
These results show that the current distribution is even, if $ \chi_l = - \chi_r $, while for the same chirality on both sides $j_y(y) $ is odd in $x$. The odd current distribution induces a magnetic field even in $x$, peaked and well localized at the junction. The transverse extension is of order of the coherence length. 

The current density is also influenced by the Josephson phase $\phi$. By comparison we find that
\begin{align}
\int_{-\infty}^0 dx\; j_y^{({\rm odd})}(x) & =-4e \eta_0^2 (A_{\gamma} +B_{\gamma} \cos \phi)(\chi_l+\chi_r), \nonumber  \\[1mm]
& = \frac{c}{4 \pi} H_z(x=0,\phi) ,
\end{align}
where we used  $ \partial H_z /\partial x = 4 \pi j_y (x) /c $ with $ H_z(0) $ corresponding to the (maximal) field at the center of the junction ($x=0$). With Eqs.~(\ref{eqn:gam}) and (\ref{eqn:hz}) we then find the relation,
\begin{equation}
\frac{\gamma}{\lambda_J^2} (1-\delta_{\gamma} \cos \phi) = \frac{\pi w_J}{\Phi_0} H_z(0,\phi),
\end{equation}
which leads to
\begin{equation}
\tilde{H}_z(\phi) =  \frac{H_z(0,\phi)}{2}.
\end{equation}
$\tilde{H}_z(\phi) $ may be interpreted as the averaged field induced by the spontaneous currents along the junction. As the solution of the effective free energy shows, this field $ \tilde{H}_z $ is screened far from the edges in an extended Josephson junction. If the currents are even in $x$ for $ \chi_r = - \chi_l $, the field $ H_{z} (x) $ is an odd function and cancels when averaged such that $ \gamma =0$.

\bibliography{references}

\end{document}